%% file: main.tex
\documentclass[submission,copyright,creativecommons]{eptcs}
\providecommand{\event}{PLACES 2017} 
\usepackage{breakurl}             
\usepackage{underscore}           

\usepackage[T1]{fontenc}
\usepackage[utf8]{inputenc}

\usepackage{amsmath,amssymb,amsthm}
\usepackage{xspace}
\usepackage{tabularx}

\input{macros}

\title{Inferring Types for Parallel Programs}
\author{
Francisco Martins
\institute{LaSIGE, Faculty of Sciences, University of Lisbon}
\and
Vasco Thudichum Vasconcelos
\institute{LaSIGE, Faculty of Sciences, University of Lisbon}
\and  
 Hans Hüttel
\institute{Aalborg Universitet}
}
\def\titlerunning{Inferring Types for Parallel Programs}
\def\authorrunning{F. Martins, V.T. Vasconcelos \&  H. Hüttel} 
\def\copyrightholders{Martins, Vasconcelos \& Hüttel} 
\begin{document}
\maketitle
\input{abstract}
\input{introduction}
\input{partypes}
\input{inference}
\input{example}
\input{discussion}

\bibliographystyle{eptcs}
\bibliography{mpi-texts}


\end{document}

%% file: macros.tex
\newcommand{\logand}{\wedge}
\newcommand{\logor}{\vee}

\newcommand{\toolname}[1]{{\sc{#1}}}
\newcommand{\partypes}{\toolname{ParTypes}}

\newcounter{paranum}[subsection]

\newcommand{\keyword}[1]{\textsf{\upshape\small #1}}

\newcommand{\messagek}{\keyword{message}}

\newcommand{\intk}{\keyword{int}}

\newcommand{\rankk}{\keyword{rank}} 

\newcommand{\foreachk}{\keyword{foreach}}

\newcommand{\allreducek}{\keyword{allreduce}}

\newcommand{\skipk}{\keyword{skip}}

\newcommand{\identifier}[1]{\keyword{#1}}
\newcommand{\size}{\identifier{size}} 
\newcommand{\rank}{\identifier{rank}} 

\newcommand{\xvar}[1]{\mathit{#1}}
\newcommand{\xrefinement}[3]{\{\xvar{#1}\colon{#2}\mid{#3}\}}

\newcommand{\pmessage}[3]{\messagek\,\,{#1}\,\,{#2}\,\,{#3}}

\newcommand{\pallreduce}[2]{\allreducek\,\,\xvar{#1}\colon{#2}.\,\,}

\newcommand{\predicatename}[1]{\mathbf{#1}}

\newcommand{\dtypep}{\predicatename{dtype}}

\newcommand{\truep}{\predicatename{true}}

\newcommand{\isExp}[3][\Gamma]{{#1}\vdash{#2}:{#3}}
\newcommand{\isProg}[3][\Gamma]{{#1}\vdash{#2}:{#3}}
\newcommand{\isProc}[3][\Gamma]{{#1}\vdash{#2}:{#3}}

\newcommand{\trueProp}[2][\Gamma]{{#1}\vdash{#2}\ \truep}

\newcommand{\equalDtypes}[3][\Gamma]{{#1} \vdash  {#2}\equiv{#3}:\dtypep}

\newcommand{\equalTypes}[3][\Gamma]{{#1}\vdash{#2}\equiv{#3}}

\newcommand{\merge}[2]{{#1}  \parallel_k {#2}}

\newcommand{\ruleskip}{.5ex} 

\usepackage{listings}
\usepackage{color}

\definecolor{darkviolet}{rgb}{0.5,0,0.4}
\definecolor{darkgreen}{rgb}{0,0.4,0.2} 
\definecolor{darkblue}{rgb}{0.1,0.1,0.9}
\definecolor{darkgrey}{rgb}{0.5,0.5,0.5}
\definecolor{lightblue}{rgb}{0.4,0.4,1}

\lstdefinestyle{eclipse}{
  breaklines=true,
  basicstyle=\sffamily\Small,
  emphstyle=\color{red}\bfseries, 
  keywordstyle=\color{darkviolet}\bfseries,
  commentstyle=\color{darkgreen},
  stringstyle=\color{darkblue},
  numberstyle=\color{darkgrey},
  emphstyle=\color{red},
  morecomment=[s][\color{lightblue}]{/**}{*/},
  showstringspaces=false,
  numbers=left,
  tabsize=2
}

\lstdefinelanguage{protocol}{
  extendedchars=true,
  showstringspaces=false,
  escapechar=@,
  basicstyle=\ttfamily\small,
  morekeywords={val,size,message,broadcast,loop,choice,foreach,gather,allgather,reduce,allreduce,scatter,imessage,wait,protocol,if,else},
  morekeywords={and,or,not},
  morekeywords={float,integer,positive,natural},
  morekeywords={length,forall},
  morekeywords={max,min,sum,prod,nand,land,band,lor,bor,lxor,bxoer,minloc,maxloc},  
  sensitive=false,
  morecomment=[l]{//}, 
  morecomment=[s]{/*}{*/}, 
  morestring=[b]",
  tabsize=2
}

\lstdefinelanguage{cpt}{ 
  style=eclipse,
  extendedchars=true,
  showstringspaces=false,
  basicstyle=\ttfamily\small,
  morekeywords={val,size,send,receive,broadcast,gather,allgather,reduce,allreduce,scatter,let,in,skip,if,then,else,for,do,while},
  morekeywords={and,or,not},
  morekeywords={float,integer,positive,natural,ref},
  sensitive=false,
  morecomment=[l]{//}, 
  morecomment=[s]{/*}{*/}, 
  morestring=[b]",
  tabsize=2,
  literate={>=}{$\ge$}1{<=}{$\le$}1
}

\lstdefinelanguage{VCC}{
  language=C,
  basicstyle=\ttfamily\small, 
  extendedchars=true,
  escapechar=@,
  showstringspaces=false,
  morekeywords={bool,datatype,lambda,integer,result,ghost,reads,writes,requires,ensures,pure,axiom,forall,out,assert,assume,thread_local,thread_local_array,true},
  tabsize=2,
}

\usepackage[scaled]{beramono}
\newcommand\Small{\small}
\newcommand\RSmall{\fontsize{7.5}{8}\selectfont} 



%% file: abstract.tex
\begin{abstract}
  The Message Passing Interface (MPI) framework is widely used in
  implementing imperative programs that exhibit a high degree of
  parallelism. The \partypes{} approach proposes a behavioural type
  discipline for MPI-like programs in which a type describes the
  communication protocol followed by the entire program.  Well-typed
  programs are guaranteed to be exempt from deadlocks. In this paper
  we describe a type inference algorithm for a 
  subset of the original system; the algorithm allows to statically
  extract a type for an MPI program from its source code.
\end{abstract}


%% file: introduction.tex
\section{Introduction}

\lstset{language=protocol}

Message Passing Interface (MPI) has become generally accepted as the
standard for implementing massively parallel programs.
%
An MPI program is composed of a fixed number of processes running in
parallel, each of which bears a distinct identifier---a
\emph{rank}---and an independent memory. Process behaviour may depend
on the value of the rank.
Processes call MPI primitives in order to communicate. Different forms
of communication are available to processes, including
\emph{point-to-point} message exchanges and \emph{collective}
operators such as broadcast.

Parallel programs use the primitives provided by MPI by issuing calls
to a dedicated application program interface. As such the level of
verification that can be performed at compile time is limited to that
supported by the host language. Programs that compile flawlessly can
easily stumble into different sorts of errors, that may or not may be
caught at runtime. Errors include processes that exchange data of
unexpected types or lengths, and processes that enter deadlocked
situations.
The state of the art on the verification of MPI programs can only
address this challenge partially: techniques based on runtime
verification are as good as the data the programs are run with;
strategies based on model checking are effective only in verifying
programs with a very limited number of processes. We refer the reader
to Gopalakrishnan et al.~\cite{formal-verification-mpi} for a discussion
on the existing approaches to the verification of MPI programs.

\partypes{} is a type-based methodology for the analysis of C programs
that use MPI
primitives~\cite{Lopez:2015:PVM:2814270.2814302,vasconcelos.etal:deductive-verification-of-mpi-protocols}. Under
this approach, a type describes the protocol to be followed by some
program. Types include constructors for point-to-point messages, e.g.
\lstinline|message from to float[]|, and constructors for collective
operations, e.g. \lstinline|allreduce min integer|. Types can be further
composed via sequential composition and primitive recursion, an
example being 
\lstinline|foreach i: 1..9 message 0 i|.
\emph{Datatypes} describe values exchanged in messages and in
collective operations, and include \lstinline|integer| and
\lstinline|float|, as well as support for arrays \lstinline|float[]|
and for refinement types that equip types with refinement conditions,
an example being \lstinline@{v:integer|v>0}@.
Index-dependent types allow for protocols to depend on values
exchanged in messages; an example of this is \lstinline@allreduce min x:{v:integer|1<=v<=9}.message 0 x@.
Our notion of refinement types is inspired by Xi and Pfenning
~\cite{xi.pfenning:dependent-types-practical-programming}, where
datatypes are restricted by indices drawn from a decidable domain.

The idea of describing a protocol by means of a type is inspired by
multiparty session types (MPST), introduced by Honda et
al.~\cite{DBLP:journals/jacm/HondaYC16}.
MPST feature a notion of global types describing, from a all-inclusive
point of view, the interactions all processes engage upon. A
projection operation extracts from a global type the local type of
each individual participant.
\partypes{} departs from MPST in that it does not distinguish between
local and global types. Instead the notion of types is equipped with a
flexible equivalence relation.  Projection can be recovered by type
equivalence in the presence of knowledge about process ranks, e.g.,
\lstinline@rank:{x:integer|x=2}@ $\vdash$
\lstinline@message 0 1 integer@ $\equiv$
\lstinline@skip@, where \lstinline@skip@ describes the empty interaction.

The type equivalence relation is at the basis of our strategy for type
reconstruction:
\begin{itemize}
\item We analyse the source code for each individual process,
  extracting (inferring) for each process a type that governs that
  individual process;
\item We then gradually merge the thus obtained types, while
  maintaining type equivalence.
\end{itemize}
This approach is related to that of Carbone and Montesi
\cite{DBLP:journals/corr/abs-1302-6331}, where several choreographies
are merged into a single choreography, and to the work of Lange and
Scalas \cite{DBLP:journals/corr/LangeS13} where a global type is
constructed from a collection of contracts.

Typable programs are assured to behave as prescribed by the type,
exchanging messages and engaging in collective operations as detailed
in the type. Moreover, programs that can be typed are assured to be
deadlock free~\cite{Lopez:2015:PVM:2814270.2814302}.
As such, programs that would otherwise deadlock cannot be typed,
implying that the inference procedure will fail in such cases,
rendering the program untypable.



%% file: partypes.tex
\section{The $n$-body pipeline and its type}

\lstset{language=C}
\input{nbodypipe-c}

We base our presentation on a classical problem on parallel
programming.
The $n$-body pipeline computes the trajectories of $n$ bodies that
influence each other through gravitational forces.
The algorithm computes the forces between all pairs of bodies,
applying a pipeline technique to distribute and balance the work on a
parallel architecture. It then determines the bodies'
positions~\cite{brinch:1991}.

The program in Figure~\ref{fig:running-example} implements this
algorithm.
Each body (henceforth called particle) is represented by a quadruple
of \lstinline|floats| consisting of a 3D position and a mass.
The program starts by connecting to the MPI middleware (line 15), and
then obtains the number of available processes and its own process
number, which it stores in variables \lstinline|size| and
\lstinline|rank| (lines 16--17).
The overall idea of the program is as follows: (a) each process starts
by obtaining a portion of the total number of particles,
\lstinline|MAX_PARTICLES|, and computes the trajectories (line~19).
Then, (b) each process enters a loop that computes
\lstinline|NUM_ITER| discrete steps. In each iteration (c) the
algorithm computes the forces between all pairs of particles. It
accomplishes this in two phases: (c.1) compute the forces among its
own particles (lines~22--23), and (c.2) compute the forces between
its particles and those from the neighbour processes (lines~25--36).
Towards this end, each process passes particles to the right process
and receives new particles from the left (lines 26--32). Then it
compute the forces against the particles received (line 33--34). After
\lstinline|size-1| steps all processes have visited all particles.
Then, (d) each process computes the position of its particles
(line~37), which results in the computation of a local time
differential (\lstinline|dt_local|), and (e) updates the simulation
time (\lstinline|sim_t|).

The simulation time is incremented by the minimum of the local time
differentials of all processes. In order to obtain this value, each
process calls an \lstinline|MPI_Allreduce| operation (line~38). This
collective operation takes the contribution of each individual process
(\lstinline|dt_local|), computes its minimum (\lstinline|MPI_MIN|),
and distributes it to all processes (\lstinline|dt|).
The minimum is then added to the simulation time (line~39).
The program terminates by disconnecting from the MPI middleware 
(line~41).

Communication is performed on a ring communication topology. The
conditional statement within the loop (lines~26--32) breaks the
communication circularity. Because operations \lstinline|MPI_Send| and
\lstinline|MPI_Recv| implement \emph{synchronous} message passing, a
completely symmetrical solution would lead to a deadlock with all
processes trying to send messages and no process ready to receive.

From this discussion it should be easy to see that the communication
behaviour of 3-body pipeline can be described by the protocol (or
type) in Figure~\ref{fig:n-body-protocol}.
The rest of this abstract describes a method to infer the type in
Figure~\ref{fig:n-body-protocol} from the source code in
Figure~\ref{fig:running-example}.


%% file: nbodypipe-c.tex
\begin{figure}[t!]
{\footnotesize
\begin{lstlisting}[language=VCC,basicstyle=\RSmall\ttfamily,numbers=left,breaklines=true,xrightmargin=2.5em]
#define MAX_PARTICLES 10000
#define NUM_ITER      5000000

void InitParticles(float* part, float* vel, int npart);
float ComputeForces(float* part,  float* other_part, float* vel, int npart);
float ComputeNewPos(float* part, float* pv, int npart, float);

int main(int argc,char** argv) {
  int  rank, size, iter, pipe, i;
  float sim_t, dt, dt_local, max_f, max_f_seg;
  float particles[MAX_PARTICLES * 4];   /* Particles on all nodes */
  float pv[MAX_PARTICLES * 6];   /* Particle velocity */
  float send_parts[MAX_PARTICLES * 4], recv_parts[MAX_PARTICLES * 4]; /* Particles from other processes */
    
  MPI_Init(&argc, &argv);
  MPI_Comm_rank(MPI_COMM_WORLD, &rank);
  MPI_Comm_size(MPI_COMM_WORLD, &size);

  InitParticles(particles, pv, MAX_PARTICLES / size);
  sim_t = 0.0f;
  for (iter = 1; iter <= NUM_ITER; iter++) {
    max_f_seg = ComputeForces(particles, particles, pv, MAX_PARTICLES / size);
    memcpy(send_parts, particles, MAX_PARTICLES / size * 4);
    if (max_f_seg > max_f) max_f = max_f_seg;
    for (pipe = 0; pipe < size - 1; pipe++) {
      if (rank == 0) {
        MPI_Send(send_parts, MAX_PARTICLES / size * 4, MPI_FLOAT, rank == size - 1 ?  0 : rank + 1, ...);
        MPI_Recv(recv_parts, MAX_PARTICLES / size * 4, MPI_FLOAT, rank == 0 ? size - 1 : rank -1,  ...);
      } else {
        MPI_Recv(recv_parts, MAX_PARTICLES / size * 4, MPI_FLOAT, rank == 0 ? size - 1 : rank -1,  ...);
        MPI_Send(send_parts, MAX_PARTICLES / size * 4, MPI_FLOAT, rank == size - 1 ?  0 : rank + 1, ...);
      }
      max_f_seg = ComputeForces(particles, recv_parts, pv, MAX_PARTICLES / size);
      if (max_f_seg > max_f) max_f = max_f_seg;
      memcpy(send_parts, recv_parts, MAX_PARTICLES / size * 4);     
    }
    dt_local = ComputeNewPos(particles, pv, MAX_PARTICLES / size, max_f);
    MPI_Allreduce(&dt, &dt_local, 1, MPI_FLOAT, MPI_MIN, ...);
    sim_t += dt;
  }
  MPI_Finalize();
  return 0;
}
\end{lstlisting}
\caption{Excerpt of an MPI program for the n-body pipeline problem
 (adapted from~\cite{Gropp:1999:UMP:330577})}
\label{fig:running-example}
}
\end{figure}


%% file: inference.tex
\section{The problem of type inference}

Given a parallel program $P$ composed of $n$ processes (or
expressions) $e_0, \dots, e_{n-1}$, we would like to find a common
type that types each process $e_i$, or else to decide there is no such
type.
We assume that \size{} is the only free variable in processes, so that
the typing context only needs an entry for this variable. We are then
interested in a context where \size{} is equal to~$n$, which we write
as $\size\colon\{x\colon\intk\mid x=n\}$ and abbreviate to~$\Gamma^n$.
Our type inference problem is then to find a type $T$ such that
$\isExp[\Gamma^n]{e_i}{T}$, or else decide that there is no such type.

\input{nbodypipe-prot}


We propose approaching the problem in two steps:
\begin{enumerate}
\item From the source code $e_i$ of each individual process extract a
  type $T_i$ such that $\isExp[\Gamma^n]{e_i}{T_i}$;
\item From types $T_0,\dots,T_{n-1}$ look for a type~$T$ that is equal
  to all such types, that is, $\equalTypes[\Gamma^n]{T_i}{T}$.
\end{enumerate}

Then, from these two results, we conclude that
$\isExp[\Gamma^n]{e_i}{T}$, hence that $\isProg[\Gamma^n]{P}{T}$, as
required.

We approach the \emph{first step} in a fairly standard way:
\begin{itemize}
\item Given an expression $e_i$, collect a system of equations
  $\mathcal D_i$ over datatypes and a type $U_i$;
\item Solve $\mathcal D_i$ to obtain a substitution $\sigma_i$. We then
  have $\isProc[\Gamma^n]{e_i}{U_i\sigma_i}$, as required for the first
  phase.
  If there is no such substitution, then $e_i$ is not typable.
\end{itemize}

For this step we introduce variables over datatypes. Then we visit the
syntax tree of each process and, guided by the typing
rules~\cite{Lopez:2015:PVM:2814270.2814302}, collect restrictions (in
the form of a set of equations over datatypes) and a type for the
expression. We need rules for expressions, index terms (the arithmetic
in types), and propositions.  We omit the rules for extracting a
system of equations and a type from a given expression.
Based on the works by Vazou et al.\ \cite{DBLP:conf/esop/VazouRJ13}
and Rondon et al.\ \cite{DBLP:conf/pldi/RondonKJ08}, we expect the
problem of solving a system of datatype equations to be decidable.

We address the second step in more detail. The goal is to build a type
$T$ from types $T_0,\dots,T_{n-1}$. We start by selecting some type
$T_i$ and merge it with some other type $T_j$ (for $i\neq j$) to
obtain a new type. The thus obtained type is then merged with another
type~$T_k$ ($k \neq j,i$), and so forth. The result of merging
all the types is the sought type~$T$. The original inference problem
has no solution if one of the merge operations fail.


%% file: nbodypipe-prot.tex
\begin{figure}[t!]
  \begin{lstlisting}[language=protocol,xleftmargin=2.5em,numbers=left]
  foreach iter: 1..5000000
    foreach pipe: 1..2
        message 0 1 float[1000000 / 3 * 4];
        message 1 2 float[1000000 / 3 * 4];
        message 2 0 float[1000000 / 3 * 4]
   allreduce min float
  \end{lstlisting}
  \caption{Protocol for the parallel n-body algorithm with three processes}
  \label{fig:n-body-protocol}
\end{figure}


%% file: example.tex
\section{Merging types}
\label{sec:example}

\lstset{language=protocol}

\input{merge-def}

We give an intuitive overview of the merge operation, discuss its
rules and apply them to our running example.
The intuition behind the merge operator is the following: 
\begin{itemize}
\item messages must be matched exactly once by the sender and the
  receiver processes (the two endpoints of the communication);
\item collective operations (\lstinline|allreduce|, for example)
  establish horizontal synchronisation lines among all processes,
  meaning that all processes must perform all communications
  (collective or not) before the synchronisation line, carry out the
  collective operation, and then proceed with the remainder of the
  protocol.
\end{itemize}
Having this in mind, the merge rules make sure that collective
operations match each other and that messages are paired together
before and after each collective operation.

The merge operation receives a typing context $\Gamma$, the type
merged so far $T$, the type to be merged~$U$ and its rank $k$, to
yield a new type $V$. We write all this as follows
$\Gamma \vdash \merge T U \leadsto V$. The typing context contains
entries for variables \size{} and \rank, the latter recording the
ranks whose types have been merged.
This context will then be updated with new entries arising from
collective (dependently typed) operations, such as $\allreducek$. An
excerpt of rules defining the merge operation is in
Figure~\ref{tab:merge}.

We first discuss merging $\skipk$ and $\messagek$ types.  There are
ten different cases that we group into the five categories detailed
below.
Notice that a $\pmessage {i_1} {i_2} {D_1}$ appearing as the left
operand of a merge is equivalent to $\skipk$ when both $i_1$ and $i_2$
are different from all ranks merged so far, which we write as
$i_1, i_2 \neq \rankk$. Otherwise, when $i_1 = \rankk$ or
$i_2 = \rankk$, the message is the endpoint of a communication between
ranks $i_1$ and $i_2$ that are already merged.
When $\pmessage {i_3} {i_4} {D_2}$ appears as the right operand of a
merge at rank $k$ it is equivalent to $\skipk$ when both $i_3$ or
$i_4$ are not $k$, which we abbreviate as $i_3, i_4 \neq
k$.
Otherwise, when $i_3 = k$ or $i_4 = k$, the message is the endpoint of
a communication with rank $k$.
Rule names try to capture these concepts. For instance, rule skip-msgS
merges $\skipk$ (left operand) with a message (right operand) that is
semantically equivalent to $\skipk$, whereas rule skip-msg designates
the merging of $\skipk$ with a message that is not equivalent to
$\skipk$.
We proceed by analysing each category.
\begin{description}
\item[merge yields $\skipk$.] In this case both operands are
  semantically equivalent to $\skipk$. This category comprises rules
  skip-skip, skip-msgS, msgS-skip (not shown), and msgS-msgS. We
  include the appropriate premises for enforcing that one or both
  parameters are equivalent to $\skipk$, depending on the message
  being the left or the right operand. For instance, rule skip-skip
  has no premises, while rule msgS-msgS includes two premises to make
  sure that both messages are equivalent to $\skipk$.
\item[merge yields the left operand.] In this category the left
  operand is not equivalent to $\skipk$, whereas the right operand
  is. It encompasses rules msg-skip and msg-msgS (not shown).
  Apart from the condition enforcing that the left message is not
  equivalent to $\skipk$ ($i_1 = \rankk \logor i_2 = rank$), rank $k$
  being merged must not be the source or the target of the
  message. Would this be the case and the program has a deadlock,
  since the messages on the left talk about rank $k$ (either as a
  source or a target) and the type at rank $k$ is $\skipk$ (or
  equivalent to it), meaning that the merged messages will never be
  matched.
\item[merge yields the right operand.] In this case the left operand
  is semantically equivalent to $\skipk$, and the right operand is
  not. The category includes rules skip-msg and msgS-msg (not
  shown). The message is from or targeted at rank $k$
  ($i_3 = k \logor i_4 = k$). We also need to check that the other
  rank of the message (the source or target that is different from
  $k$) is still to be merged ($i_3, i_4 \neq \rankk$). Why?  Because
  otherwise the type of the other endpoint is already merged and is
  $\skipk$ (the left operand), therefore the message at rank $k$ (the
  right operand, which is not $\skipk$) is never going to be matched,
  indicating the program has a deadlock.
\item[messages are the endpoints of the same communication.] In this
  category (rule msg-msg-eq) the messages correspond to the two
  endpoints of a communication. The result of the merge is the left
  operand, which is semantically equivalent to the right one. No
  message is semantically equivalent to $\skipk$ as witnessed by the
  premises.
  Additionally 
  we need to check that the source and the target ranks, as well as
  the payload, of the two messages coincide.
\item[messages are the endpoints of different communications.]  This
  last category includes messages that are the endpoints of two
  different communications. The result of the merge is an interleaving
  of the messages. The messages are semantically different from
  $\skipk$ and are unrelated. The category includes rules msg-msg-left
  (not shown) and msg-msg-right. As in the previous category we check
  that no message is semantically equivalent to $\skipk$.
  Additionally, we check that the messages do not interfere, that is,
  that their ranks are not related. These two rules can be
  non-deterministically applied in an appropriate way to match the
  types. 
\end{description}

There are no rules to merge messages against collective operations,
since this is not admissible; the merging of messages against foreach
loops is left for future work. Collective operations can only be
merged against each other (cf.\ rule allred-allred). We omit the
rules for other MPI collective operations for they follow a similar
schema.
In this paper we only merge foreach loops against foreach loops. Refer
to the next section for a discussion about the challenges on this
subject.

The last three rules apply to the sequential composition of types:
rule seq-seq allows for types to be split at the sequential operator
($;$) and merged separately; rules msgT-msgT-left and msgT-msgT-right
(not shown) allow for the non-deterministic ordering of unrelated
messages, as described for rules msg-msg-left and msg-msg-right, but
here at the level of the sequential composition of types.
The last rule allows for messages after the last collective
communication (if any) to be merged.
For the sake of brevity, we also omit rules for the sequential
composition of skip types.

We now outline how merging works on our running example.
%
%
Fix \lstinline|size = 3|. From the program in
Figure~\ref{fig:running-example} extract \lstinline|size| programs,
one per rank, in such a way that programs do not mention variable
\lstinline|rank|. We leave this to the reader.

Run the first step of our procedure on each program to obtain the
three types below, where \lstinline|D| is the datatype
\lstinline|float[MAX_PARTICLES / size * 4]|.

\begin{center}
  \begin{minipage}{.32\textwidth}
    For \lstinline|rank 0|:
\begin{lstlisting}
 foreach iter: 1..5000000
   foreach pipe: 1..2
     message 0 1 D;
     message 2 0 D
 allreduce min float
\end{lstlisting}
  \end{minipage}
  \begin{minipage}{.32\textwidth}
    For \lstinline|rank 1|:
\begin{lstlisting}
 foreach iter: 1..5000000
   foreach pipe: 1..2
     message 0 1 D;
     message 1 2 D
 allreduce min float
\end{lstlisting}
  \end{minipage}
  \begin{minipage}{.32\textwidth}
    For \lstinline|rank 2|:
\begin{lstlisting}
 foreach iter: 1..5000000
   foreach pipe: 1..2
     message 1 2 D;
     message 2 0 D
 allreduce min float
\end{lstlisting}
  \end{minipage}
\end{center}

Run the second step as follows. We only show the merging of the
various messages; the cases of \lstinline|foreach| and
\lstinline|allreduce| are of simple application.

We start by taking the type for the process at rank 0 and merge it
with that of rank 1.  The initial typing context $\Delta_1$ says that
the type on the left corresponds to rank 0 in a total of 3, which we
write as
$\size \colon \xrefinement x \intk {x = 3}, \rankk\colon \xrefinement
x \intk {x = 0}$.
Using rules seq-seq, msg-msg-eq, and msg-msg-right we have:
\begin{center}
  $\Delta_1 \vdash\;\;$
  \begin{minipage}{.37\textwidth}
\begin{lstlisting}
message 0 1 D; || message 0 1 D;
message 2 0 D  || message 1 2 D
                 1
\end{lstlisting}
  \end{minipage}
  $\quad\leadsto\quad$
  \begin{minipage}{.20\textwidth}
\begin{lstlisting}
message 0 1 D;
message 1 2 D;
message 2 0 D
\end{lstlisting}
  \end{minipage}
\end{center}

Then we merge the resulting type with that of rank 2. This time we
need a typing context $\Delta_2$ that records the fact that the type
on the left corresponds to ranks 0 and 1. We write it as 
$\size \colon \xrefinement x \intk {x = 3}, \rankk\colon \xrefinement
x \intk {x = 0 \logor x = 1}$.
Using rules msgT-msgT-left, seq-seq, msg-msg-eq (x2), we get:

\begin{center}
  $\Delta_2 \vdash\;\;$
  \begin{minipage}{.37\textwidth}
\begin{lstlisting}
message 0 1 D; || message 1 2 D;
message 1 2 D; || message 2 0 D
message 2 0 D    2
\end{lstlisting}
  \end{minipage}
  $\quad\leadsto\quad$
  \begin{minipage}{.20\textwidth}
\begin{lstlisting}
message 0 1 D;
message 1 2 D;
message 2 0 D
\end{lstlisting}
  \end{minipage}
\end{center}

The type obtained is that of Figure~\ref{fig:n-body-protocol}.













%% file: merge-def.tex
\begin{figure}[t!]
  \centering
  \begin{gather}
    \tag{skip-skip}  
                     \Gamma \vdash 
                     \merge{\skipk}{\skipk}
                     \leadsto \skipk
    \\[\ruleskip]
    \tag{skip-msgS} \frac{
                        \trueProp{
                          i_3, i_4 \neq k}
                      }{
                      \Gamma \vdash 
                      \merge {\skipk}{\pmessage {i_3} {i_4} D}
                      \leadsto \skipk
                     }
    \\[\ruleskip]
    \tag{msgS-msgS}  \frac{
                       \trueProp{i_1, i_2 \neq \rankk \logand
                       i_3, i_4 \neq k}
                     }{
                     \Gamma \vdash 
                     \merge{\pmessage {i_1} {i_2} {D_1}}{\pmessage {i_3} {i_4} {D_2}}
                     \leadsto \skipk
                     } 
    \\[\ruleskip]
   \tag{msg-skip} \frac{
                        \trueProp{(i_1 = \rankk \logor i_2 = \rankk)
                          \logand i_1, i_2 \neq k}
                      }{
                      \Gamma \vdash 
                      \merge {\pmessage {i_1} {i_2} D_1}{\skipk}  
                      \leadsto \pmessage {i_1} {i_2} D_1
                     }
    \\[\ruleskip]
   \tag{skip-msg} \frac{
                        \trueProp{i_3, i_4 \neq \rankk \logand 
                          (i_3 = k \logor i_4 = k)}
                      }{
                      \Gamma \vdash 
                      \merge {\skipk} {\pmessage {i_3} {i_4} D_2} 
                      \leadsto \pmessage {i_3} {i_4} D_2
                     }
    \\[\ruleskip]
    \tag{msg-msg-eq}  \frac{
                       \trueProp{(i_1 = \rankk \logor i_2 = \rankk) \logand
                       (i_3 = k \logor i_4 = k) \logand i_1 = i_3
                       \logand i_2 = i_4}
                     \qquad
                     \equalDtypes{D_1}{D_2}
                     }{
                     \Gamma \vdash 
                     \merge{\pmessage {i_1} {i_2} {D_1}}{\pmessage {i_3} {i_4} {D_2}}
                     \leadsto \pmessage {i_1} {i_2} {D_1}
                     } 
    \\[\ruleskip]
   \tag{msg-msg-right} \frac{
                        \trueProp{(i_1 = \rankk \logor i_2 = \rankk)
                          \logand
                          (i_3 = k \logor i_4 = k) \logand 
                          i_1 \neq i_4 \logand i_2 \neq  i_3
                        }
                      }{
                     \Gamma \vdash 
                     \merge{\pmessage {i_1} {i_2} {D_1}}
                     {\pmessage {i_3} {i_4} {D_2}}
                     \leadsto \pmessage {i_3} {i_4} {D_2}; \pmessage {i_1} {i_2} {D_1}
                     }
    \\[\ruleskip]
   \tag{allred-allred} \frac{
                       \equalDtypes{D_1}{D_2} 
                       \qquad
                       \Gamma, x\colon D_1 \vdash \merge {T_1} {T_2}
                       \leadsto T_3
                      }{
                       \Gamma \vdash 
                       \merge {\pallreduce x {D_1} T_1} 
                       {\pallreduce x {D_2} T_2} 
                       \leadsto 
                       \pallreduce x {D_1} T_3
                     }\\[\ruleskip]
   \tag{foreach-foreach} \frac{
                        \trueProp{i_1 = i_2 \logand i_1' = i_2'}
                       \qquad
                       \Gamma, x\colon \xrefinement y \intk {
                         i_1 \leq y \leq i_1'} \vdash \merge {T_1} {T_2}
                       \leadsto T_3
                      }{
                       \Gamma \vdash 
                       \merge {\foreachk\; x\colon i_1..i_1'.T_1} 
                       {\foreachk\; x\colon i_2..i_2'.T_2} 
                       \leadsto 
                       \foreachk\; x\colon i_1..i_1'.T_3
                     }\\[\ruleskip]
   \tag{seq-seq} \frac{
                       \Gamma \vdash \merge {T_1}{T_3}
                       \leadsto T_5
                       \qquad
                       \Gamma \vdash \merge {T_2}{T_4}
                       \leadsto T_6
                      }{
                     \Gamma \vdash 
                     \merge{T_1;T_2}{T_3;T_4}
                     \leadsto T_5;T_6
                     }
    \\[\ruleskip]
   \tag{msgT-msgT-left} \frac{
                        \trueProp{(i_1 = \rankk \logor i_2 = \rankk)
                        \logand (i_3 = k \logor i_4 = k) 
                      \logand i_1 \neq i_4 \logand i_2 \neq  i_3}
                        \quad
                       \Gamma \vdash \merge {T_1} {\pmessage {i_3} {i_4} {D_2}; T_2}
                       \leadsto T_3
                      }{
                     \Gamma \vdash 
                     \merge{\pmessage {i_1} {i_2} {D_1};
                       T_1}{\pmessage {i_3} {i_4} {D_2}; T_2}
                     \leadsto \pmessage {i_1} {i_2} {D_1}; T_3
                     }
    \\[\ruleskip]
   \tag{skip-msgT} \frac{
                       \Gamma \vdash \merge {\skipk} {\pmessage {i_3} {i_4} {D}}
                       \leadsto T_2
                        \quad
                       \Gamma \vdash \merge {\skipk} {T_1}
                       \leadsto T_3
                      }{
                     \Gamma \vdash 
                     \merge{\skipk}{\pmessage {i_4} {i_4} {D}; T_1}
                     \leadsto T_2; T_3
                     }
  \end{gather}
  \caption{Rules defining the merge partial function (excerpt)}
  \label{tab:merge}
\end{figure}


%% file: discussion.tex
\section{Discussion}
\label{sec:discussion}

\lstset{language=protocol}

The procedure outlined in this paper is not complete with respect to
the \partypes{} type system~\cite{Lopez:2015:PVM:2814270.2814302}. We
discuss some of its shortcomings.

\paragraph{Variables in MPI primitives}

In order to increase legibility, code that sends messages to the left
or to the right process in a ring topology often declares variables
for the effect. The original source code~\cite{Gropp:1999:UMP:330577}
declares a variable \lstinline[language=C]|right| with value
\lstinline[language=C]|rank == size - 1 ?  0 : rank + 1|. The
\lstinline[language=C]|MPI_Send|
operation in line~27 is then written as follows:
\begin{lstlisting}[language=C]
MPI_Send(sendbuf, MAX_PARTICLES / size * 4, MPI_FLOAT, right, ...);
\end{lstlisting}
In this particular case the value of \lstinline|right| is computed
from the two distinguished \partypes{} variables---\size{} and
\rank---and it may not be too difficult to replace
\lstinline[language=C]|right| by
\lstinline[language=C]|rank == size - 1 ? 0 : rank + 1| in the type.
In general, however, the value of variables such as
\lstinline[language=C]|right| may be the result of arbitrarily complex
computations, thus complicating type inference in step one of our
approach.
In addition, indices present in types can only rely on variables whose
value is guaranteed to be uniform across all processes. It may not be
simple to decide whether an index falls in this category or not.

\paragraph{Parametric types}

The type in Figure~\ref{fig:n-body-protocol} fixes the number of
bodies in the simulation (line~1). The original source code, however,
reads this value from the command line using
\lstinline[language=C]|atoi(argv[1])|. The \partypes{} language
includes a dependent product constructor \lstinline|val| that allows
to describe exactly this sort of behaviour:
\begin{lstlisting}
val n: natural.
foreach iter: 1..5000000
  foreach pipe: 1..2
    message 0 1 float[n / 3 * 4]
    ...
\end{lstlisting}
The \partypes{} verification procedure seeks the help of the user in
order to link the value of expression
\lstinline[language=C]|atoi(argv[1])| in the source code to variable
\lstinline|n| in the
type~\cite{Lopez:2015:PVM:2814270.2814302,vasconcelos.etal:deductive-verification-of-mpi-protocols}.
When we think of type inference, it may not be obvious how to resolve
this connection during the first step of our proposal.

\paragraph{Type inference and type equivalence}

\partypes{} comes equipped with a rich type theory, allowing in
particular to write the three messages in the protocol
(Figure~\ref{fig:n-body-protocol}, lines 3--5) in a more compact form:
\begin{lstlisting}
foreach i: 0..2
  message i (i == 2 ? 0 : i + 1) float[n / size * 4]
\end{lstlisting}
It is not clear how to compute the more common \lstinline|foreach|
protocol from the three messages, but this intensional type is not
only more compact but also conductive of  further generalisations
of the procedure, as outlined in the next example.


\paragraph{The number of processes is in general not fixed}

A distinctive feature of \partypes---one that takes it apart from all
other approaches to verify MPI-like code---is that verification does
not depend on the number of processes. The approach proposed in this
paper, however, requires a fixed number of processes, each running a
different source code (all of which can nevertheless be obtained
from a common source code, such as that in
Figure~\ref{fig:running-example}). Then, the first step computes one
type per process, and the second step merges all these types into a single
type.
The \partypes{} verification procedure allows to check the program in
Figure~\ref{fig:running-example} against a protocol for an arbitrary
number of processes (greater than 1), where the internal loop (lines
2--5) can be written as
\begin{lstlisting}
foreach pipe: 1..size-1
  foreach i: 0..size-1 
    message i (i + 1 < size ? i + 1 : 0) float[n / size * 4]
\end{lstlisting}
The merge algorithm outlined in this paper crucially relies on a fixed
number of types, one per process, and is not clear to us how to
relieve this constraint.

\paragraph{One-to-all loops}

The type presented in the paragraph above contains two
\lstinline|foreach| loops: the former corresponds to an actual loop in
the source code (lines 23--33), the latter to a conditional (lines
26--32).
By expanding the source code in Figure~\ref{fig:running-example} for
each different process rank, the first step of our proposal extracts
types of the same ``shape'' for all processes, as we have seen in
Section~\ref{sec:example}.
Now consider the following code snippet, where process 0 sends a
message to all other processes:
\begin{lstlisting}[language=C]
if (rank == 0)
  for(i = 1; i < size; i++)
    MPI_Send(sendbuf, n / size * 4, MPI_FLOAT, i, ...);
else
  MPI_Recv(recvbuf, n / size * 4, MPI_FLOAT, 0,  ...);
\end{lstlisting}
Fixing \lstinline|size == 3| as before, the first phase yields the following
types:
\\[1ex]
\begin{tabular}[h]{ll}
  \lstinline|foreach i: 1..2 message 0 i float[n * 4]|
  &
    for rank 0,
  \\
  \lstinline|message 0 1 float[n * 4]|
  &
    for rank 1, and
  \\
  \lstinline|message 0 2 float[n * 4]|
  &
    for rank 2.
\end{tabular}
\\[1ex]\noindent leaving for phase two the difficult problem of
merging one \lstinline|foreach| type against a series of
\lstinline|message|
types. When the limits of the \lstinline|foreach| loop are constant, we can
unfold it and merge the thus obtained sequence of messages as in
Section~\ref{sec:example}, but this is, in general, not the case.




